\documentclass[12pt, preprint]{aastex}
\usepackage{pdfsync}
\usepackage{natbib}
\DeclareMathSymbol{\varOmega}{\mathord}{letters}{"0A}
\DeclareMathSymbol{\varSigma}{\mathord}{letters}{"06}
\DeclareMathSymbol{\varPsi}{\mathord}{letters}{"09}

\title{Roadmap For Small Bodies Exploration: Theoretical Studies}
\author{D. Nesvorn\'y$^1$, A. Youdin$^2$, S. E. Dodson-Robinson$^3$, A. Barr$^1$, E. Asphaug$^4$, D. J. Scheeres$^5$}
\affil{(1) Department of Space Studies, Southwest Research Institute}
\affil{(2) Harvard-Smithsonian Center for Astrophysics}
\affil{(3) Department of Astronomy, University of Texas at Austin}
\affil{(4) Department of Earth \& Planetary Sciences, University of California at Santa Cruz}
\affil{(5) Department of Aerospace Engineering Sciences, University of Colorado at Boulder}

\begin{document}
\maketitle
\vspace*{1.cm}
This document covers five broad topics from theory of small bodies: planetesimal formation, cosmochemistry, thermal 
evolution, collisions and dynamics. Each of these topics is described in a separate section, where we prioritize 
selected main issues over completeness. The text points out the principal unresolved problems in each area, 
and suggests ways how progress can be made. This includes support for new code development, observations 
and experimental work that can be used to constrain theory, new research directions, and studies of cross-over  
regimes where the system's behavior is determined by several competing processes. The suggested development areas 
are placed in the context of NASA space exploration. 

\section{Planetesimal Formation}
A primary rationale for small bodies research is to improve our understanding of how planetesimals form --- a vital 
step in the growth of terrestrial and habitable planets.  To maximize the scientific return from exploration missions, 
it is imperative to support theoretical work on planetesimal formation.  It is also beneficial to adapt mission 
designs to address pressing theoretical issues, when feasible and well-justified.  As a concrete example, since equal 
mass Kuiper belt binaries are a sensitive probe of formation models, a New Horizons encounter with such a system is 
well-motivated.

Broadly speaking there are two ways to improve theories of planetesimal formation.  The first is to improve our 
understanding of the fundamental physical processes at play.  The second is to fit observed properties of small 
bodies to a phenomenological model by varying uncertain parameters.  These approaches are not mutually exclusive, 
but in reality compromises must be made.  Studies of physical processes may lack all the necessary ingredients --- 
or computational resources --- to match reality.  Models built to match data may include physically unjustified 
assumptions. Pursuing studies that correctly weight both approaches will lead to the greatest scientific progress.

\subsection{Formation Theories}
The textbook theory that planetesimals formed by orderly collisional growth is becoming increasingly questioned.  
Both experiments and theoretical arguments have highlighted the low efficiency of sticking in the millimeter to 
meters size range.  This has coincided with a burst in progress in dynamical formation mechanisms to trigger 
planetesimal formation by gravitational collapse, as described below.  However these dynamical models only intensify 
the need to understand collision outcomes.  The extent to which collisional growth proceeds towards centimeter sizes, 
or possibly beyond, remains a key uncertain input to the dynamical models.  For larger planetesimals, however they 
may form, ongoing collisional evolution (see \S4) must be included for comparison with observations.

The initial version of the gravitational hypothesis was proposed by Safronov (1969) and Goldreich \& Ward 
(1973). The critique by Weidenschilling (1984) of the model significantly advanced the study of nebular dynamics.  
He argued that vertical shear instabilities in the disk midplane would trigger turbulence that halts the vertical 
sedimentation of particles to the midplane.  Consequently, densities should be too low for direct gravitational 
collapse.  Subsequent work, notably by Cuzzi, Dobrovolskis \& Champney (1993) confirmed this obstacle.  
The seriousness of the obstacles to both gravitational collapse and collisional growth left planetesimal 
formation theory in a difficult position.

\begin{figure}[t!]
\epsscale{0.5}
\plotone{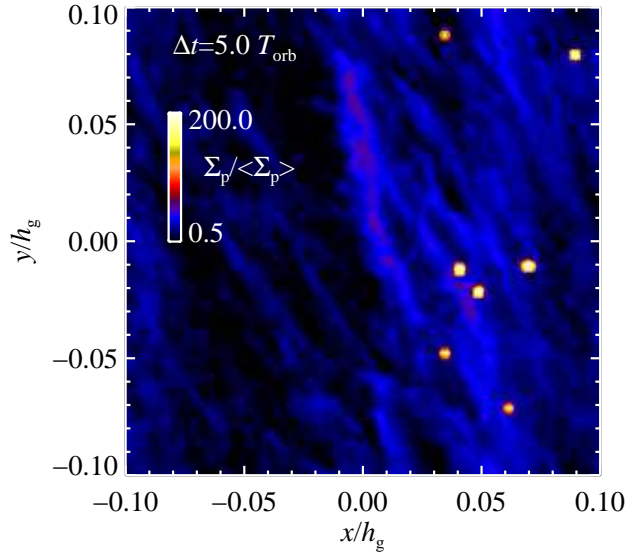}
\caption{Particle column density, $\Sigma_{\rm p}$, showing the formation of seven gravitationally bound clumps in 
a 3D, vertically stratified, shearing box simulation of unmagnetized gas and superparticles with stopping time 
0.1--0.4 (Johansen, Youdin \& MacLow 2009). The initial box-averaged particle-to-gas density 
$\langle \Sigma_{\rm p} \rangle/ \langle \Sigma_{\rm g} \rangle = 0.02$.  
The $x (y)$ axis is parallel to the radial (azimuthal) direction, and measured in units of the gas scale height.  
The simulation first evolved for 40 orbits without self-gravity, during which particles settled to the midplane and 
triggered vertical shearing and streaming instabilities; strong clumping resulted.  This snapshot is taken 5 orbits 
after self-gravity was turned on. The bound fragments contain $\sim$20\% of the total mass in solids; each has a mass 
comparable to a compact planetesimal having a size 100--200 km. Figure from Johansen, Youdin \& MacLow (2009).}
\label{youdin_fig}
\end{figure}

Advances in dynamical studies now offer possible resolutions.  Sekiya (1998) and Youdin \& Shu (2002) showed how 
vertical  shear instabilities were weakened when the ratio of solids-to-gas was enhanced above Solar abundance values, 
offering a route to gravitational collapse.  This mechanism is supported by the finding that extrasolar planets are 
more abundant around high metallicity stars, which presumably had dustier protostellar disks.  However the issue 
remained for how to enhance the abundance of solids relative to gas in disks around Solar-type stars.  

Many mechanisms have been proposed for the aerodynamic concentration of solids in a gas disk.  These include gaseous 
spiral arms, persistent vortices, photoevaporation of gas, pileups of solids due to radially varying drift rates.  
Cuzzi et al. (2001) proposed that turbulence, typically thought of as a stirring agent, could also concentrate solids 
on the scale of eddies due to centrifugal expulsion. The mechanism has advantage of efficiently concentrating 
mm-sized solids for reasonable assumptions about the turbulence.  By contrast the other mechanisms mentioned above 
work best on larger, meter-sized solids.  The size difference reflects the aerodynamic coupling or stopping timescale.  
For meter sizes, the stopping time is just orbital timescale, which is the most relevant for most dynamical 
processes. However fast-swirling eddies interact with smaller solids that have a shorter stopping time.

The streaming instability (SI) is a powerful and ``active" clumping mechanism that arises spontaneously due to drag 
forces in disks.  By contrast the above-mentioned clumping mechanisms are ``passive" since particles respond to an 
assumed flow of the gas.   While discovered analytically by Youdin \& Goodman (2005), numerical simulations of the 
SI by Johansen, Youdin and collaborators (Johansen \& Youdin 2007; Johansen, Youdin \& MacLow 2009) 
revealed that SI clumping amplitudes are large enough to trigger gravitational collapse (Fig. \ref{youdin_fig}).  
The strongest clumping amplitudes are seen for 
$\sim$10 cm solids.  While smaller solids appear to clump less efficiently --- as reinforced by the simulations of 
Bai \& Stone (2010) --- numerical difficulties with small stopping times are still a factor.  One hope is that the 
small scale clumping described by Cuzzi et al. (2001) could interact constructively with SI, but this is still 
difficult to simulate directly.

A surprising finding of recent work is that planetesimals which form by gravitational collapse could be many 
hundreds of kilometers in radius.  This is much larger than the canonical value of several kilometers from 
Goldreich \& Ward (1973) because collapse begins at much larger densities due to aerodynamic clumping.
Johansen et al. (2007) formed 1000~km radius equivalent (since the collapse did not proceed to solid 
densities) bodies in simulations with with SI clumping and forced (by magnetic fields) turbulence.  Johansen, 
Youdin \& MacLow (2009) found 100-200 km (equivalent) radius planetesimals from SI clumping of slightly smaller 
solids and no forced turbulence, i.e. only turbulence driven by particle-gas interactions. A more complete 
survey of parameter space will help understand these dependencies and lead to predictions of initial size 
distributions.  

Cuzzi, Hogan \& Shariff (2008) argued that the small scale clumping by forced turbulence could also extend to 
larger scales and produce 100 km scale planetesimals.  While this conjecture is not amenable to direct simulation, 
the theoretical basis of the scaling arguments can be developed. 

\subsection{Discriminant Tests of Formation Theories}  
Different planetesimal formation theories should have different implications for the properties of small body 
populations. For example, the gravitational collapse has the potential to quickly create objects with 
characteristic large size, while two-body collisions may lead to more gradual growth and result in a more 
continuous size spectrum. Progress toward discriminating between different formation theories can therefore be made by 
carefully comparing the theoretical predictions with properties of asteroids and Kuiper belt objects (KBOs). 
Unfortunately, except of the recent work by Schlichting \& Sari (2011), who estimated the initial size function 
(ISF) of KBOs for the standard growth by two-body collisions, detailed theoretical predictions are not generally
available (see also Cuzzi et al. 2010).   

While it may be possible to derive ISF of planetesimals from analytic scaling arguments, the analytical work is 
more likely to be valuable elsewhere, for example when identifying (and characterizing) the principal physical 
processes that contribute to planetesimal formation. Reliable IMF predictions will instead be most likely 
obtained from numerical simulations that are probably more amenable to incorporating the complex physics of 
non-linear solid-gas coupling. Presently, however, the predictive power of simulations is still strongly limited 
by large CPU requirements that stem from need to have both the detailed resolution and global coverage of disk 
processes. Support should thus be given to the development of new, more efficient numerical codes that will be 
able to speed up calculations and allow scientists to more fully explore parameter space.  

The theoretical suggestion that planetesimals could form big very quickly found support in Morbidelli et al. 
(2009), who argued that the size distribution of the largest asteroids is likely imprinted by the collapse 
process.  This conclusion arises from the difficulty of explaining the size distribution of large asteroids 
with standard collisional growth. While suggestive, this work also illustrates the difficulty in discriminating 
between formation and evolution processes. This is because the ISF can be modified by disruptive collisions 
and dynamical depletion processes occurring since planetesimal formation 4.6 Gy ago. For example, it is believed 
that most asteroids with radius $R<10$ km are fragments of disruptive collisions (Bottke et al. 2005), while the population
of large asteroids may have been depleted by a factor 10-1000 by dynamical processes (Petit et al. 2002). 
Understanding the evolution processes is thus an important issue that arises when it is attempted to derive 
ISF from observations, and support should be given to those who seek funding for such a research. 

While KBO populations also show characteristic size distributions that can arise from the formation process, 
the evolution history of the KBOs is significantly more uncertain, making interpretation difficult. 
Specifically, it is not well understood whether the cold classical KBOs, a dynamical component between 
42-48 AU having low eccentricities and low inclinations, formed in situ at 42-48 AU or was implanted in that 
region from elsewhere (Levison et al. 2008). The size distribution shape of cold classical KBOs with $R>10$ 
km is similar to that of the main belt asteroids, showing a break at $R=50$ km. The break can suggest ISF 
with a preferred size of planetesimals with $R=50$ km (Nesvorn\'y et al. 2010a, 2011), or can be a signature of 
disruptive collisions that depleted the population of objects with $R<50$ km (Pan \& Sari 2005). This issue is of 
crucial importance for the planetesimal formation theories in the outer solar system. 

\section{Cosmochemistry}
We focus on Ceres in this section because it is one of the nearby objects that can potentially show interesting 
chemistry that is relevant to planet formation. Moreover, with the Dawn mission, we have the real 
opportunity to use Ceres as a chemistry lab and calibrate our cosmochemical models.  

Ceres, the largest asteroid in our Solar System, is a keystone for astronomers trying to reconstruct the growth 
histories of planets.  A member of asteroid class C (carbonaceous), its spectrum shows evidence of organic 
compounds on its surface. Furthermore, Ceres bulk density of only 2.08 g cm$^{-3}$ (Thomas et al. 2005) indicates that 
it contains up to 30\% water ice by mass. The abundance of both water ice and organics in Ceres provides evidence 
that it formed in a cold part of the Solar Nebula --the disk of raw planet-forming material that surrounded the 
young Sun-- where volatile compounds such as water and ammonia could freeze. Since each ``ice'' has a different 
characteristic temperature at which it clings to planet-building dust particles, measurements of the compositions 
of icy asteroids contain a fossil record of the planet-forming environment.

Although astronomers have predicted the mass fraction of ices in Uranus and Neptune (Hubbard et al. 1995, 
Dodson-Robinson \& Bodenheimer 2010), most of the heavy elements in gas giants are locked deep in their interiors, 
invisible to observers. By contrast, C-class asteroids such as Ceres are ideal proving grounds for theories 
of planet formation because they can hold ices on their surfaces. NASA's Dawn spacecraft, with its Gamma Ray and 
Neutron Detector (GRaND), is already en route to the asteroid belt and will reach Ceres in 2015 (Rayman et al. 2006). 
GRaND will map out the atomic composition of Ceres' surface, allowing astronomers and geophysicists to determine 
Ceres' ice inventory and temperature distribution and reconstruct its evolutionary history.

Chemical evolution models can maximize the potential of the Dawn Mission by computing the abundances and 
distributions of ices in Ceres from initial freezeout through asteroid growth and cooling. By determining the 
range of possible surface compositions for Ceres today, cosmochemical analysis will allow the GRaND instrument to 
make the leap from present-day composition to planetary archeology: its data will help us determine the correct 
asteroid formation pathway among many possibilities. 

\begin{figure}[h!]
\epsscale{0.8}
\plotone{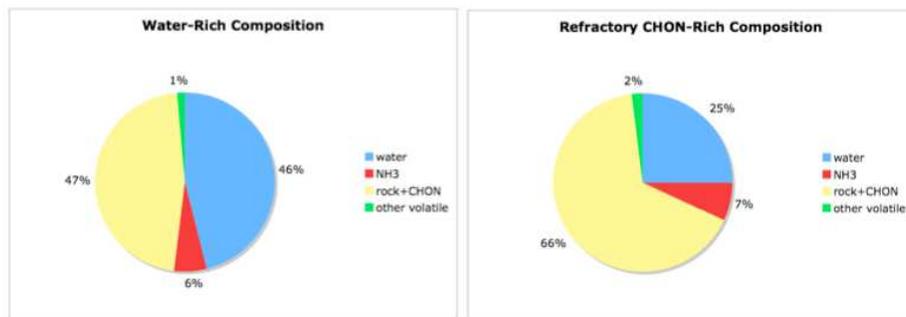}
\caption{Ceres' current composition may have evolved from an initially water-rich chemical inventory (left), where 
water was lost due to internal heating. It is also possible that water loss over time was negligible and Ceres' 
initial and present-day compositions are similar (right). The difference in initial water abundance between the 
two models pictured is due to different assumptions about the sequestration of carbon in solid (refractory) grains 
versus lightweight, volatile hydrocarbons. GRaND measurements may distinguish between these two models.}
\label{sally_fig}
\end{figure}

Following is a brief list of possible formation scenarios and associated observables for Ceres, based on the solar 
nebula models of Dodson-Robinson et al. (2009) and the asteroid geophysical models of Castillo-Rogez \& McCord (2010):

\begin{enumerate}
\item Although pure ammonia has an extremely low freezing point, ammonia-water mixtures may deposit on dust 
surfaces at much higher temperatures. Ceres may contain up to 7\% ammonia by mass. (Fig. \ref{sally_fig}). 
The presence of 
ammoniated minerals would suggest that water and ammonia formed a mixed ice matrix during Ceres' formation. 
Incorporation of ammonia into water ice would allow subsurface ocean formation on small bodies more distant from 
the Sun, as ammonia lowers the melting point of water. Ganymede, Callisto, Titan and Enceladus are all predicted 
to have subsurface oceans (Spohn \& Schubert 2003, Hussmann et al. 2006, Lopes et al. 2007).

\item Ceres accreted almost 50\% water ice by mass (Fig. \ref{sally_fig}). A key question is about the fraction 
of that water: bound to minerals (e.g., hydrated silicates and salts) and that amount available to form an icy shell.  
The large-scale migration of 
water through the asteroid might have created hydrated minerals on Ceres' surface that are observable today. Ceres 
may even be experiencing volatile loss today, similar to the outgassing from asteroid Lutetia observed by the ROSINA 
experiment aboard Rosetta (Wurz et al. 2010).
\item Although much of the organic material on Ceres' surface is likely graphite and kerogen, some of it may be 
hydrocarbons such as acetylene and ethane. Hydrocarbons on Ceres' surface would provide strong evidence that Ceres' 
formation zone was colder than 60~K.
\end{enumerate}


An inventory of the organic material on Ceres' surface is especially important in light of the recent discovery 
that Allende's remnant magnetization was most likely imprinted by the convecting metallic core of a partially 
differentiated parent body (Weiss et al. 2010). Elkins-Tanton et al. (2010) argue that chondritic meteorites are 
not remnants of pristine bodies but are instead parts of the unmelted mantles of partially differentiated asteroids. 
Indeed, ongoing analyses of Rosetta observations will reveal whether (21) Lutetia has the high density and CV-like 
surface that characterize the disrupted CV parent body. Ceres and Pallas both have shapes and bulk densities 
consistent with partial differentiation.

Volatile retention implies late formation relative to the formation of the Sun and the CAIs for two reasons: (1) 
the inner solar nebula must cool enough for ices to freeze, and (2) $^{26}$Al decay leads to volatile loss. However, 
partially differentiated asteroids are thought to be members of an early generation of planetesimals. A differentiated 
yet volatile-rich asteroid such as Ceres may point to an extremely brief planetesimal formation epoch, at least 
in the inner solar system. Castillo-Rogez \& McCord (2010) have already begun pinpointing Ceres' formation time 
relative to the CAIs. Ceres shape data, gravitational field measurements and surface composition inventory from 
Dawn will therefore be critical to exploiting Ceres' potential as a planet formation chronometer.

\section{Thermal Evolution Models}
\begin{figure}[t!]
\epsscale{0.5}
\plotone{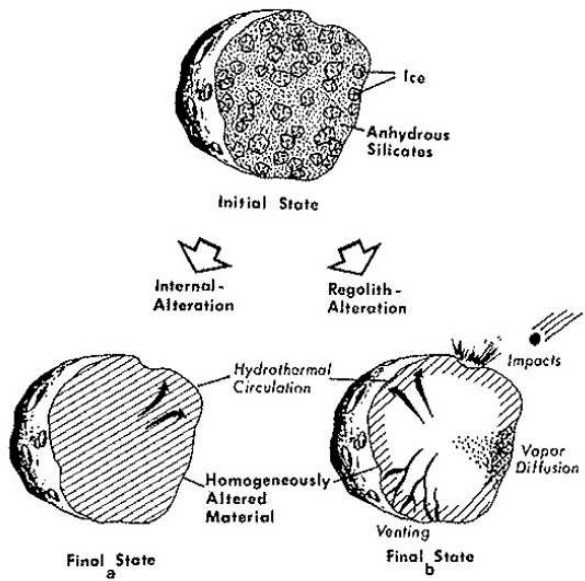}
\caption{Schematic illustration of the processes that shape the interiors of small bodies, focused on the CI/CM parent body (Grimm \& McSween 1989),
including hydrothermal alteration of silicates in the body's interior, hydrothermal circulation driven by interior temperature gradients, vapor diffusion
in the body's interior, venting through surface cracks, and impacts.}
\label{grimm_mcsween}
\end{figure}

Traditionally, thermal evolution models of small bodies are divided into three sub-classes: 
\begin{enumerate}
\item Early thermal evolution of a hypothesized CI/CM parent body (see Figure \ref{grimm_mcsween}; radius $R=50$ to 100 km) composed of ice 
and rock to determine conditions under which silicates in their interiors are hydrothermally altered, with 
the goal of deducing what the mineralogy of CI and CM meteorites tells us about conditions in the early 
solar system (e.g., Grimm \& McSween 1989, Travis \& Schubert 2005).
\item Early thermal evolution of rock/metal bodies such as Vesta ($R\sim265$ km), focusing on chemical and physical 
differentiation and petrogenesis, with the goal of understanding the composition and mineralogy of HED 
meteorites (e.g., Gosh \& McSween 1998).
\item Thermal evolution of comets ($R\sim5$ km) due to insolation, radiogenic heating, formation and breakdown of 
clathrates, and ice phase changes in present-day conditions with the goal of understanding the timing, duration, 
and magnitude of outgassing events (see Prialnik 2000 for a comprehensive summary).
\end{enumerate}

\begin{figure}[t!]
\epsscale{0.8}
\plotone{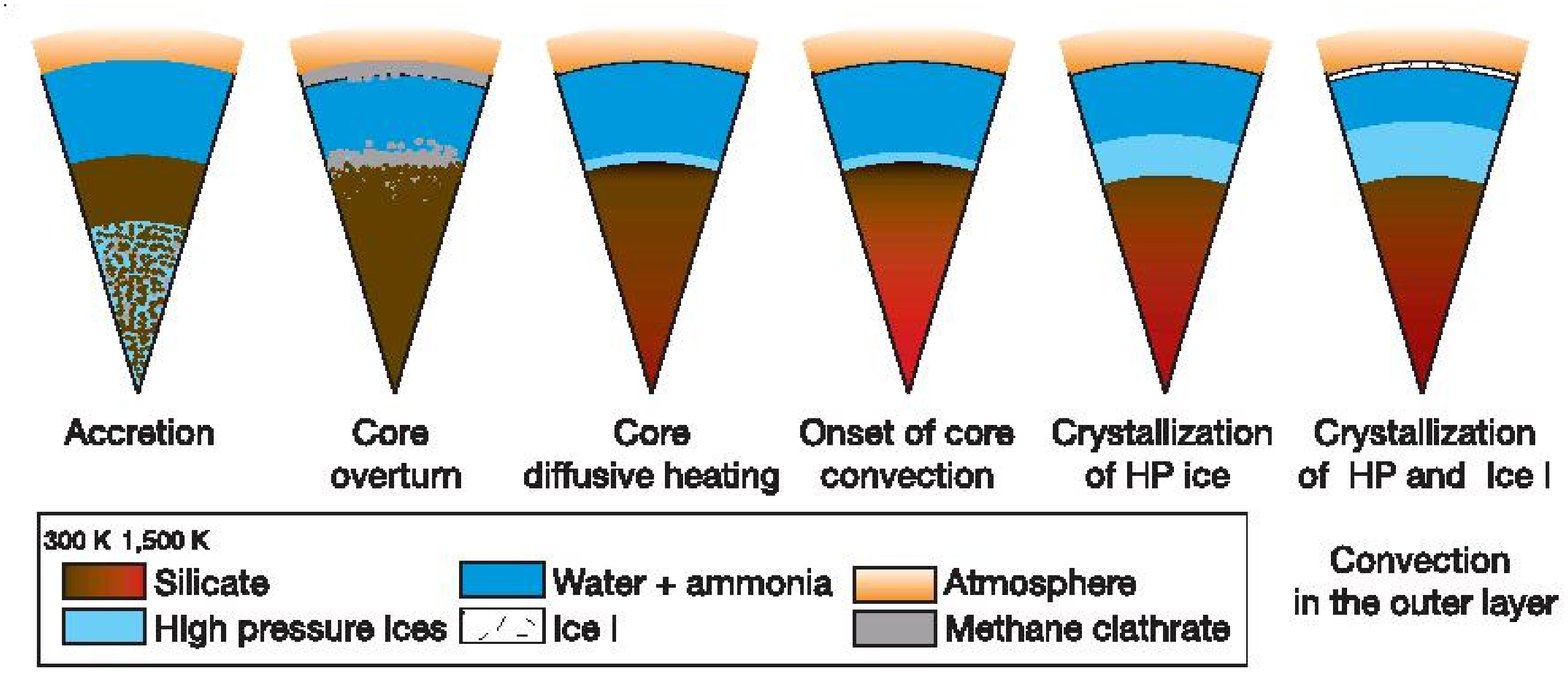}
\caption{Global thermal processes at work inside a large icy satellite, in this case, Titan (Tobie et al., 2006).  After accretion of a cold mixed ice/rock core
overlain by a layer of pure rock, the deep interior of the satellite is gravitationally unstable, and an initial core overturn is triggered (left).  The core
is heated from within by decay of long-lived radioisotopes and eventually convection begins in the core (middle).  As the satellite cools, layers of high-pressure
ice phases and ice I form at the top of the core and surface of the satellite, respectively.  Unlike the ``small body model'' in Figure \ref{grimm_mcsween}, 
hydrothermal circulation, vapor diffusion, and other processes that advect mass through interior voids and cracks are ignored.}
\label{thermal}
\end{figure}

In such models, ice/rock bodies have a period of early activity driven by accretional heating and decay of short-lived 
radioisotopes (SLRI; $^{26}$Al, $^{60}$Fe).  
These early heat sources drive differentiation of the primordial ice/rock mixture 
into a consolidated central rock core.  The rock core heats by short-lived radiogenic heating, and later by $^{40}$K 
heating, and may dehydrate, liberating even more energy.  Heat transport in the overlaying ice mantle occurs by 
solid-state conduction, hydrothermal circulation and/or diffusion of water vapor through mantle pore spaces.  The 
formation and evolution of a porous insulating regolith at the surface plays a key role in raising the interior 
temperatures of small bodies. The evolution of rock/metal bodies such as Vesta also experience a period of early 
activity driven by SLRI heating, which leads to differentiation, melting and crustal formation.  

Small body models are distinguished from thermal evolution models of solid planets and planetary satellites by 
their inclusion of heat and mass transport through diffusion or circulation of fluid or vapor through interior cracks 
or void spaces.  In large solid planets or satellites, these processes are thought to be important only in the outer 
few kilometers where lithostatic pressures are too low to close cracks and voids.  

Small body models typically focus on a single burst of activity early in the body's history, immediately after 
formation.  By contrast, thermal evolution models of solid planets and satellites typically focus on heat transfer 
over billion-year time scales and assume that the system has forgotten its initial condition (appropriate for a 
diffusive/convective planetary interior; Fig. \ref{thermal}).  

However, the line between small bodies and large planetary objects has been blurred by, for example, the 
discovery of the water-rich composition of Ceres ($R=475$ km), and geological diversity of Saturn's small ice/rock 
moons (radii range between 200 to 750 km).  In particular, Enceladus ($R=252$ km) has a geologically active region at 
its south pole driven by tidal heating (Schubert et al., 2007; Nimmo et al., 2007) characterized by large heat fluxes 
and ongoing degassing from its interior (Porco et al., 2006).  Outermost Iapetus ($R=764$ km) has a peculiar shape 
indicative of an early period of rapid rotation, and a 20-km-high equatorial ridge.  Discovery of the incomplete 
differentiation of Callisto ($R=2410$ km; Anderson et al. 1998) and Rhea (Anderson \& Schubert 2007; Iess et al., 2007) 
suggests that SLRI and accretional heating in these objects was limited (Barr \& Canup 2008), providing clues about 
the timing and duration of their formation.

Accordingly and perhaps appropriately, the boundary between thermal evolution models of small bodies and planetary 
satellites has begun to blur.  Processes typically reserved for small bodies may play a role in the geochemical 
and geophysical evolution of planetary satellites, and vice versa.  

As an example, one model proposed for the plume activity on Enceladus suggests that the observed abundances of water 
and other species in the plume can be explained by clathrate degassing (Kieffer et al., 2006). Solid-state convection, 
typically thought to occur exclusively in large planetary objects, may play a role in driving activity on Enceladus 
(Roberts \& Nimmo 2008; Barr 2008).  SLRI heating and mantle porosity have been envoked to explain the early thermal 
evolution of Iapetus (Castillo-Rogez et al., 2007). 

A removal of the artificial boundary between small body and planetary satellite models is well justified by 
recent spacecraft observations. Improved numerical methods, e.g., coupled models of hydrothermal and thermal convection 
(e.g., Travis \& Schubert 2005) can help remove the boundary between these fields.  

However, uncertainties about the material properties of ice and rock at planetary conditions have led to uncertainties 
about which physical processes occur inside a given body.  For example, solid-state convection is difficult to start, 
but can occur in the 100 kilometer-thick ice mantle of Enceladus (Barr \& McKinnon 2007). McCord and Sotin (2005) also
demonstrated that convection onset is possible in the case of Ceres. 

Castillo-Rogez and McCord (2010) pointed out
the complexity of convection modeling in the case of Ceres, because of the abundant amount of impurities expected in the early
ocean as a consequence of intense hydrogeochemistry promoted by SLRI decay. A large amount of impurities is expected
to affect the structure of the shell. Kargel (1991) showed that thick layers of hydrated salts that precipitated at the base 
of an icy shell (asteroid or icy satellite) may be subject to cryovolcanism. This was modeled in more detail by 
Prieto-Ballesteros and Kargel (2005) in the case of Europa. 


Thus, laboratory experiments must be encouraged and supported alongside new theoretical model development.  In 
particular, measurements of the porosity, thermal conductivity, yield strength, and viscosity of pure ice and ice/rock 
mixtures at confining pressures appropriate for small- to mid-sized ice/rock bodies are urgently needed.  These will 
give modeling efforts a new level of realism and modelers confidence in their results.

\section{Impact Studies}
Asteroids, comets and small satellites evolve physically in response to cratering impacts and more energetic 
collisions. The small bodies we see today are, for the most part, collisional disruption remnants derived from 
the ancestral populations of bodies that formed 4.6 billion years ago, whose characteristics and sizes remain 
poorly known, and which formed by accretionary processes involving slower collisions (as discussed in \S1). 
Meteorites from small bodies are delivered by the thousands to Earth by more recent impacts. 

Understanding impacts and collisions is fundamental to understanding small bodies, and five approaches are taken: 
\begin{enumerate}
\item simulations based on physical integrations (computer models)
\item theoretical approaches based on conservation laws and scaling
\item experimental studies in impact chambers, centrifuges, quarries, parabolic flights and drop towers, 
and (soon) sub-orbital laboratory flights
\item direct observation (natural collisions, and space missions like Deep Impact)
\item physical and petrological studies of meteorites
\end{enumerate}

\subsection{Simulations}  
Computer simulation is a rapidly advancing modern tool that unites theory and experimentation. Here it is most crucial 
to benchmark these physics-based algorithms to lab experiments, analytic solutions, and astronomical and spacecraft 
observations.  It is not enough to conserve mass, momentum and energy, and it is not sufficient to prove (say)
 that the 
method is 2nd order accurate.  Computer models of small body impacts and collisions are applied to untested domains  
and as with any software unexpected errors much be expected until the behavior in the relevant 
problem domain is fully understood.  Most notoriously, the equation of state (EOS) can meddle with the integration by 
doing unanticipated things to the sound speed, or to the internal energy.  Or, the application of a low-density cutoff 
or a sound speed minimum can suppress shocks.  High resolution may be required to capture important aspects of the 
physics and in 3D may not be achievable in near future.

To the extent that computer simulation is sometimes called the `third branch of science' benchmarking cannot be 
emphasized enough.  The strange environment of small body impacts (the nonintuitive dominance of gravity; the Coriolis 
forces; the influence of aggregate cohesion; the trapping of seismic energy) makes it difficult to decide whether or 
not a simulation has produced reasonable results, or has run to late enough time.  But once properly benchmarked, a 
simulation allows us to conduct numerical experiments that lead to understandings of planetary processes at scales far 
beyond what is attainable in the laboratory, and in regimes far too complex for theoretical analysis alone.  In the 
coming decade we can make very substantial progress if we pay attention to the pitfalls of numerical modeling.

A number of primary computer codes are used for simulations of small body impacts and collisions.  Continuum codes 
include smoothed particle hydrodynamics (SPH; Benz \& Asphaug 1995, Jutzi et al. 2008)
and various grid based codes such as CTH
(e.g. Housen \& Holsapple 2003, Leinhardt \& Stewart 2009).
These codes often include 
relatively crude rheological models (strength, friction) which may not adequately capture microgravity behavior 
especially at the very low strain rates and stresses governing the end-game of global-scale cratering 
(Asphaug \& Melosh 1993).  
Another 
limitation is computer power, where high resolution 3D runs are feasible, or runs to late time, but seldom both.  
The post-impact flow field may require an hour or more to evolve in response to a collision (the gravity timescale), 
whereas the sound-crossing time on a small body can be less than a second.  This discrepancy means that millions of 
model timesteps (a resolution element divided by the sound speed) may be needed to simulate an impact or collision to 
late time, easily tying up a modest computer facility for days, and producing unwieldy data sets.  The bigger the 
impact event, the easier it is to model from a numerical point of view, because the sound crossing time becomes 
comparable to the collision time; this creates a situation where impacts into small bodies must strive to obtain 
time on massive supercomputers, a problem similar in scope to the challenge to obtain time observing diminutive 
asteroids on large facility telescopes.

Particle codes are also applied to model rigid bodies such as spheres or collections of polyhedra, applicable at 
events that are slow enough to allow for rigid-body treatments of their constituents 
(e.g. Richardson et al. 2002, Korycansky \& Asphaug 2006). Action is instantaneous so 
these models run quickly.  Sometimes particle codes are included in hybrid schemes where a continuum approach is 
used to evolve the early stages of an impact, and a granular $N$-body code is used to evolve the late stages 
(e.g. Michel et al. 2001).  Here 
the cutting edge is the influx of new techniques adopted from the granular physics community, where the rigid body 
approach is being supplanted by elastic-dashpot and van der Waals forces, and the achievement of very high resolution 
simulations that can model a rubble pile out of realistically small grain sizes to reveal aggregate phenomena.

Continuum and particle codes have been applied to study asteroid catastrophic disruption and family formation, 
formation of binary systems, hit-and-run collisions, evolution and loss of regolith, and impact shock alteration.  
Often the simulations are at the cutting edge of what is achievable.  Looking to the advent of very inexpensive 
supercomputing on GPUs (graphics cards) in the coming decade, we shall soon be able to trace the detailed impact 
response in 3D targets with realistic compositions and structures, to late time.  While laboratory benchmarking 
is essential, meteorites serve as their own benchmarks.  For instance, impact modeling can be tied to meteorite 
constraints such as the shock/degassing levels of ordinary chondrites and other types of meteorites.  As we push to 
higher and higher resolution, treatments at the continuum level (sub-grid-scale porosity, fracture damage and mixed 
phases) will evolve into treatment at the explicit level (real voids or zones of damage in the grid, and zones of 
different composition).  But the path from here to there is not straightforward.  An order of magnitude increase 
in computer performance means a factor of $\sim$2 increase in 3D model resolution, so by Moore's law every 10 fold 
increase in 3D resolution requires $\sim$15 years.  Moreover, to date every leap in resolution has given a slightly 
different answer to a number of the more detailed modeling problems.  Computer models guide and extend our intuition 
but are not yet, except in a few well-posed domains, a replacement for direct observation and theory.

\subsection{Theoretical Approaches}
The scaling of crater dimensions applies to asteroids as well as it does to the surfaces of planets, although for 
large craters vector $\mathbf{g}$ is not nearly a constant $g_z$.  On most small bodies seen to date, a crater diameter 
is achieved that is comparable in diameter to the target, and no good theory exists to account for craters in this 
size regime.  The global impact response is very poorly understood, in varying gravity and finite geometry where the 
surface curvature is comparable to the crater diameter, and where trapped vibrations can modify the crater even as 
it is forming.  A fundamental problem is that useful scaling models assume that either strength or gravity (but not 
both) dominate the collisional response; data suggest that this is not the case for small bodies, where gravity is 
minuscule yet where so much of the post-impact morpohology looks gravity-controlled, and where strength may be 
`hiding' in the form of granular cohesion (Scheeres et al. 2010). 

Accretion involves collisions between objects that are comparable in size, at velocities comparable to or slower than 
their mutual escape velocity.  At the small end, early on, these may have involved $\sim$10 km bodies crashing together 
at meters per second; at the large end, the so-called late stage, it is believed that Mars sized planets collided 
with proto-Earth.  
Smaller-scale accretionary collisions occur at the speed of a car crash and involving 
undifferentiated materials -- a quite unstudied physics.  

The path forward is to extend, and perhaps look beyond traditional scaling approaches, by examining the results of 
new observations and simulations, perhaps adapting theoretical models from companion sciences such as seismology and 
agranular physics. 

\subsection{Experimental Studies}  
The near future promises inexpensive access to low-Earth orbit, with research-project class suborbital flights 
providing 15 minutes or more of clean microgravity.  These relatively low-cost flights will allow microgravity 
impact (and blast analog) experiments at cm- to possibly m-scale, providing data in a controlled environment that 
is likely to be closely relevant to the 100-m to km-scale events on asteroids, both those occurring naturally and 
those being triggered by artificial means.  

At a small NEO, a subspacecraft ``pod'' with total mass  less than 10 kg might be expected to deliver enough blast 
efficiency to produce a $\sim$100 m diameter crater on an asteroid the size of Itokawa ($\sim$300-500 m diameter), in a 
medium-class NASA mission where an orbiter monitors the event.  Crater formation would take hours.  The comparable 
event can be studied (if gravity-scaled) in a suborbital flight, but this would also take hours.  Because ballistic 
and explosion studies do not mix well with human spaceflight, a robotic orbital research platform for NEO studies 
would be effective for conducting experiments bridging the strength and gravity regimes. 

\begin{figure}[t!]
\epsscale{0.5}
\plotone{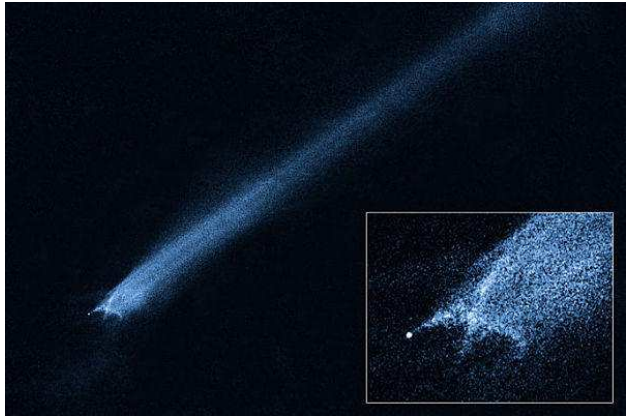}
\caption{Hubble Space Telescope observations of inner-belt asteroid P/2010 A2 show a peculiar comet-like 
morphology. The data reveal a nucleus of diameter $\approx$120 meters with an associated tail of millimetre-sized dust particles.
It is most probably a remnant of asteroidal disruption in February 2009, evolving slowly under the action of solar radiation 
pressure (Jewitt et al. 2010, Snodgrass et al. 2010). Figure from Jewitt et al. (2010).}
\label{a2}
\end{figure}

\subsection{Direct Observation}
Until an impact between two small planetary bodies is directly observed, we can study the 
aftermath in various forms: the cratering record on small bodies; the asteroid families which derive from catastrophic 
disruption events; the dust-brightening observed for a few asteroids that are probably due to recent 
impact (Fig. \ref{a2}); and (if one counts tidal collisions) the gravitational disruption of comets by Jupiter.  

Great inroads have been made relating asteroid family formation to the process of catastrophic disruption 
(Michel et al. 2001, 2002, 2003, Nesvorn\'y et al. 2006, Durda et al. 2004, 2007).  
Larger 
telescopes, and more telescope time spent on small bodies, have enabled detailed spectroscopy at high time resolution 
to understand spin rates, shapes, and compositional and thermal characteristics of family members formed millions of 
years ago.  With the advancement of synoptic sky surveys such as LSST it shall become possible to witness the aftermath 
of small bodies in the days or weeks following the smaller-scale collisions that occur sporadically.

Spacecraft make detailed observations of the cratering record of small bodies, revealing global and regional responses, 
notably the seismic shaking by impact of pre-standing topography which can be used to derive the mechanical response 
to impact.  The cratering record of a number of close-approaching NEOs have also been detected by radar telescopes at 
Arecibo and Goldstone (e.g. Benner et al. 2002),
greatly leveraging the few spacecraft encounters.  Active missions allow for direct monitoring 
of the cratering process in microgravity, and successors to Deep Impact shall conduct observational campaigns from 
orbit (see \S4.3).

\subsection{Meteorites} 
Meteorites are blasted off from asteroids and (sometimes) from the Moon and Mars by impacts.  Closely related is the 
interstellar 
dust that may have formed on comets.  Meteorites formed originally in the collisional processes that created 
asteroids, and evolved on their parent bodies, showing signatures of many kinds of collisional processing.  In the 
coming decade, for progress to be made in either field - the study of meteorites, and of small body impacts and 
collisions including the formative ones - a closer kinship is required between these sciences, which have advanced 
distinctly yet, for the most part, separately over the past decade, to the point that most meteoriticists are outsiders 
to impact studies and vice-versa.  Fewer scientists `bridge the gap' between both disciplines, as was more 
feasible 20-30 years ago when both disciplines were smaller.  A regular workshop on the collisional origin and 
evolution of meteorites and their impact delivery to Earth is recommended.

\section{Dynamics}
While significant progress has been made in understanding the dynamical evolution of small bodies, there 
remain areas where progress needs to be made. This arises in part due to the interesting and unforeseen 
configurations and morphologies of small body systems that are being discovered by observational astronomers. 
These include binary and triple systems, asteroids spinning at their disruption limit, asteroid pairs, contact 
binary structures, and fast rotating complex tumblers, among others. 

\subsection{Heliocentric Orbits}
Classical studies in dynamics have mainly focused on the heliocentric orbits of small bodies and their time 
evolution. In the past few decades many fundamental problems have been solved in this realm. Examples include 
the joint analysis of the Yarkovsky effect, resonances, and secular dynamics to clearly map out how migration 
of asteroids occurs within the main belt (e.g, Vokrouhlick\'y \& Farinella 2000, Bottke et al. 2001). By 
combining these analyses researchers now have a clear view of how the inner solar system is populated with 
small bodies.

New work in this area is in tracing out the implications of solar system evolution for the current distribution 
and dynamics of small bodies in the solar system (Minton \& Malhotra 2009). The goal is to use current and predicted 
distributions of small bodies as tests for or against various theories of planet migration. A particularly 
fruitful area of exploration is the Kuiper belt (Levison et al. 2008), as the orbital structure of this population 
is not fully explained. Specifically, it is not understood whether the classical KBOs formed in situ beyond 40
AU or were scattered into that region from $<$40 AU. 

The progress in this area can be made by better characterizing the classical population from observations because 
that will help to limit the number of theoretical possibilities. For example, the large binary fraction among 
classical KBOs appears to provide an interesting constraint (Parker et al. 2010, Nesvorn\'y et al. 2011), because
these loosely bound binary systems are fragile and become easily disrupted in some migration models. 

\subsection{Rotational Evolution}
The recent detection of the YORP effect (e.g., Rubincam 2000) has paved the pathway for the deeper understanding 
of small body spin distributions (Taylor et al. 2007, Lowry et al. 2007, Kaasalainen et al. 2007). 
Research is still trying to 
understand the full implications of YORP on the long-term rotational evolution of small bodies. Initial studies 
have been made into nearly all of the topics of interest in this area, although full convergence onto a unified 
model of YORP and its implications remains to be achieved. 

Specific areas that still need a fuller theoretical resolution and understanding include a complete model
of the spin state and obliquity evolution of a small body subject to the YORP effect, especially when bodies are 
brought down into a slow rotation state. While there are several different possible end-states for small body 
rotational dynamics (Vokrouhlick\'y et al. 2003, Scheeres \& Mirrahimi 2008), the analysis of the joint 
obliquity/spin state convergence has not been carried out in a complete manner. Of special interest is whether 
a complex rotation state can significantly alter how a small body responds to the YORP effect and whether a 
body can be trapped into a slow rotation state for long periods of time (Vokrouhlick\'y et al. 2007, Cical\`o 
\& Scheeres 2010). 

Another topic that requires fuller resolution relates to the interactions between impacts and the YORP effect 
which could play an important role in the main belt. Initial investigation of this phenomenon indicates that 
significant non-linear interactions between these effects can occur (Rossi et al. 2009, 2010). The 
relative importance of these effects as a function of size has not been analyzed either, but due to the physics 
of these effects there must be a cross-over between YORP dominance and collisional dominance.

There also remains significant uncertainty in the level of detail in modeling required to successfully account for 
the YORP effect. The poster child of this is the asteroid Itokawa, which when simply evaluated should have a 
detectable change in its spin rate due to the YORP effect (Scheeres et al. 2007), yet this change has not been 
observed. Discussions have focused on more precise modeling of the thermal budget on the asteroid surface 
(Vokrouhlick\'y \& \v{C}apek 2002, Mysen 2008), more precise accounting for internal density distributions (Scheeres \& Gaskell 2008), and higher 
resolution modeling (Nesvorn\'y \& Vokrouhlick\'y 2007, Statler 2009, Breiter et al. 2009). The presumed eventual 
detection of YORP on Itokawa will provide a stringent test for the precise modeling of the YORP effect, as a 
detailed shape model exists for this body.

The recent detection of spin state changes in comets also raises the rotational dynamics of these bodies as an 
interesting topic of study (Belton \& Drahus 2007, Knight et al. 2010). New methods should be developed to enable the spin 
state propagation of cometary bodies over long time spans, building on an averaged dynamical model such as is 
found in (Neishstadt et al. 2002) and motivated by recent models of cometary outgassing (Crifo \& Rodionov 1997). This will 
enable insight into what, if any, limiting rotational behavior is expected in the presence of random outgassing 
over comet surfaces incorporating relaxation dynamics as well. A strong motivation for this is the intuitive 
involvement between rotational dynamics and comet nucleus bursting.

\subsection{Long-Term Evolution of Multiple Component Asteroid Systems}
A new topic that requires dynamical understanding is the long-term evolution of binary and multiple component 
asteroid systems in the near-Earth object and main belt populations. The binary YORP effect (BYORP) predicts that 
the lifetimes of binary asteroids are extremely short, on the order of $10^5$ years (\'{C}uk 2007, McMahon \& 
Scheeres 2010). However, this would imply an implausibly rapid formation rate for binaries, and thus alternative 
time evolutions of these bodies have been investigated. 

Current modeling efforts are concentrating on the effect that coupled rotational and 
translational motion will have on the BYORP effect (\'{C}uk \& Nesvorn\'y 2010, McMahon \& Scheeres 2010), although there has not 
been uniform agreement on the net effect of these interactions. What is needed in this area is an analysis of the 
long-term evolution of a binary system that fully accounts for all evolutionary effects that will act on an 
asteroid system over time. Individual analyses have looked at and compared components of these effects, but 
have not systematically put them all together. Such an analysis should include tidal effects (Goldreich \& Sari 2009), 
the YORP effect on the primary (Harris et al. 2009, Fahnestock \& Scheeres 2009), the BYORP effect on the 
secondary (\'{C}uk \& Burns 2005, McMahon \& Scheeres 2010), and spin-orbit coupling between the two bodies (\'{C}uk \& Nesvorn\'y 2010, McMahon \& Scheeres 2010).

\subsection{Rubble Piles with Evolving Rotational Angular Momentum}
Related to the spin evolution of small bodies and the stability of multi-component systems is the more general 
question of how rubble pile bodies evolve in response to secular or abrupt changes in their total rotational 
angular momentum. These changes can occur slowly over time due to the YORP effect, can occur over a single 
rotation period of a body due to close planetary flybys, or can occur instantaneously due to impacts.

The basic mechanics of collections of bodies resting on each other are not well understood. Important work has 
been done at the continuum level by Holsapple, indicating that strength models may play an important role in 
the ability of small bodies to spin rapidly, and supply some understanding of how such bodies might fail 
(Holsapple 2001, 2004, 2007, 2010). However, the continuum limit is a significant 
simplification at the size scale of these small bodies, where the discrete nature of interaction must be 
accounted for. 

Some rudimentary theoretical work has been done investigating the basic rules that a rubble pile system must 
follow as its angular momentum increases over time (Scheeres 2007, 2009). Similarly, initial forays into 
the modeling of such systems has also been performed (Walsh et al. 2008, Sanchez \& Scheeres 2011). However, systematic theories for and 
modeling of discrete bodies resting on each other incorporating exogenous perturbations is a challenging problem 
and will be central to the understanding of a number of fundamental problems that span different systems across 
the solar system. Specifically, the dynamics of discrete granular systems are present during the initial 
planetesimal formation phase of the solar system, when smaller components were initially precipitating and 
clumping into larger structures. 

Numerical studies of these bodies have been made using hard-sphere codes (Richardson et al. 2005), however modeling 
the low energy, contact structure between the particles forming a rubble pile is challenging given the impulsive 
nature of interactions in these codes. Modifications to these simulations are being investigated that will allow 
these codes to be extended into this regime. More recent work has developed a modeling formulation that applies 
soft-sphere models for self-gravitating bodies in contact (Sanchez \& Scheeres 2011), leveraging from the field of granular 
mechanics. The field of numerical simulation of rubble piles with changing angular momentum is still at an early 
stage of development, and its development will be important for these systems to be better understood. A specific 
challenge in this field is tracking how energy and angular momentum can be redistributed within a rubble pile as a 
function of its total angular momentum and the manner in which it changes.

\begin{figure}[t!]
\epsscale{0.5}
\plotone{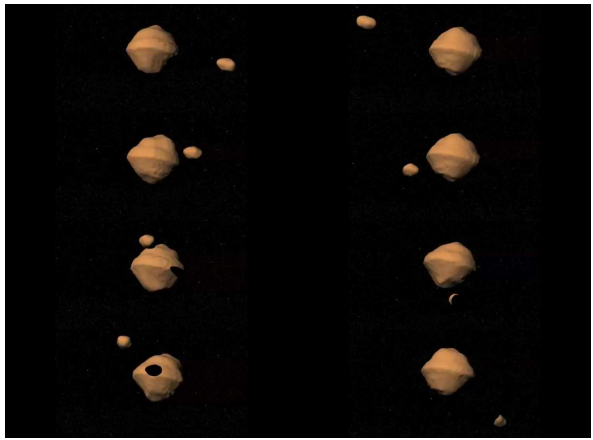}
\caption{High resolution radar images reveal near-Earth asteroid (66391) 1999 KW4 to be a binary system. 
The $\approx$1.5-km-diameter
primary is an unconsolidated gravitational agregate dominated by an equatorial ridge. The $\approx$0.5-km
secondary is elongated and probably denser than the primary. Figure from Ostro et al. (2006).}
\label{kw4}
\end{figure}

\subsection{Formation and Failure of Binary and Multi-Asteroid Systems}
The rotational fission of rubble piles has been implicated in the formation of binary systems (Fig. \ref{kw4}; Scheeres et al. 2006, 
Scheeres 2007, Walsh et al. 2008) and in the formation of asteroid pairs (Vokrouhlick\'y \& Nesvorn\'y 2008, 
Scheeres 2009, Pravec et al. 2010). The specific 
process by which a fissioned rubble pile evolves to its end state is not understood, however. Specifically, we find 
a diverse morphology of small body types equally subject to YORP yet not fully explained with one theoretical 
understanding, including binary and triple systems, contact binary systems, tumbling asteroids, reshaped asteroids, 
and other more specific morphologies. 

Recent evidence on asteroid pairs provides interesting clues to the bi-modal 
mass distribution structure of asteroids that form asteroid pairs (Pravec et al. 2010). Additionally, theoretical 
work predicts that all initially fissioned bodies to be unstable (Scheeres 2009), whether or not they are 
energetically bound to each other, although the full implications of this are still being worked out. Specifically, 
the initial evolution of two component bodies formed by fission remains unmodeled in a detailed sense, although 
simple initial models have been developed and are showing that the evolution is highly chaotic and has significant 
spin-orbit coupling occurring over the first few days, weeks and months of these systems (Jacobson \& Scheeres 2009, 
2010). How this can lead to the observed small body systems is a key topic for future 
understanding.

\subsection{Celestial Mechanics of Low Energy Systems}
At the mathematical level, there is a dearth of research into celestial mechanics systems at the low energies we 
find for rubble pile bodies. A celestial mechanics view of the rubble pile problem would involve the imposition 
of a finite density for constituent bodies, making them have finite radii and allowing them to lie on each other 
in relative equilibria in addition to orbital equilibria. A systematic study of this problem from the point of 
view of rigorous mathematics could yield significant fruit, and would also provide specific results that could be 
used to validate and verify the performance of numerical simulations. One possible motivating thesis statement 
would be to evaluate all possible stationary energy configurations of celestial mechanics systems with finite radii 
as a function of angular momentum. Initial studies have just looked at two bodies in interaction, but could be 
extended to multi-body systems (Scheeres 2007).

\subsection{Plasma and Weathering on the Surfaces of Small Bodies}
Small bodies lie in extreme environments, and are themselves extreme and exotic locales. How these bodies interact 
with their external environment and what physics are important for their own evolution and structure are essentially 
open questions.

The general interaction of small body surfaces with the solar system environment is a current field of research. 
It is well known that exposure of asteroid surfaces to sunlight and the solar wind causes changes in their spectral 
signatures over time, called weathering. There have been many advances in understanding these weathering physics in 
recent decades, however a full understanding of the process and its speed is still lacking.

In a particularly interesting recent development, correlations between Earth flybys and S to Q Type transitions among 
NEO have been discovered (Nesvorn\'y et al. 2005, 2010b; Binzel et al. 2010). However, the details of this process are 
not understood, as the initially proposed tidal interactions seem to be too weak to be responsible for mechanically 
altering the surface of an asteroid and exposing fresh material. This is motivating detailed investigations into the 
interactions of charged surfaces with a planet's magnetosphere to ascertain whether some non-gravitational effect 
is either causing resurfacing or can systematically lead to a change in the spectral signature of a body.

Related to these phenomenon is dust levitation caused by plasma fields on the surface of airless bodies and charging 
through the photoemission effect. Dust clouds were observed on the moon and current theories link them with the surface 
plasma environment and charging of dust grains, although the detailed mechanisms or the necessary surface environment 
that leads to levitation are not well understood. Similar levitation has also been hypothesized on asteroids 
(Lee 1996, Colwell et al. 2005) and invoked to explain the dust ponds on the Eros surface (Robinson et al. 2001). However, a full accounting 
of forces on an airless body places other restrictions on how a particle may be mobilized from the surface of the moon 
or an asteroid, and seems to throw some of the assumptions about how levitation physically occurs into doubt. There is 
currently no end-to-end theoretical explanation for this phenomenon, and thus this is ripe for additional research.

\subsection{Evolution of Comet Surfaces}
Recent images of comet surfaces and close-in environments have only added mystery to how their surfaces evolve and 
change over time. It is not an overstatement to say that each comet imaged at close range has had a markedly different 
surface morphology. Some theories and simple explanations of how comets work have been developed (Jewitt 2004, 
Goguen et al. 2008, Belton et al. 2008, Belton \& Melosh 2008), however these do not provide for a uniform theory of comet surfaces that 
accounts for all the different observed surface morphologies. With the imminent arrival of Rosetta at its target comet 
67P/Churyumov-Gerasimenko, there will be a wealth of detail not available before concerning the physical evolution 
and morphology of a comet's surface. Thus, development of clear theoretical approaches to this problem would be able 
to be tested or evaluated with this coming data set.

\subsection{Physical Forces in Micro-Gravity Settings}
The surfaces and interiors of small bodies can have ambient accelerations less than micro-gravity due to their small 
mass and potentially rapid spin rates. When the weights of objects become vanishingly small it is feasible that other 
physical forces and effects may play significant roles. These include self-gravitational attraction between bodies, 
van der Waals cohesion, cold welding of materials, and a variety of other effects that may not be relevant for larger 
bodies. Theoretical predictions have been made concerning these effects (Scheeres et al. 2010), but laboratory tests 
must still be made to fully evaluate how the near absence of gravity modifies the physics of interaction between 
bodies.

Of specific interest relating to this are the bulk constitutive laws appropriate for small bodies in the micro-gravity 
regime. Specifically, a better understanding of how tidal dissipation occurs in these systems and how interiors of 
small bodies will strain in response to stress fields are required. Classical theories tend to be related to deformation 
of a continuous media. However, for small bodies tribology effects must be included and considered in order to develop 
a full model of these interactions. Initial work in this area has been done (Holsapple 2010, Goldreich \& Sari 2009), yet a 
specific grounds-up physical analysis of this effect must still be made.


\clearpage

\begin{thebibliography}

\bibitem{and2} Anderson, J. D., Schubert, G., 2007, Saturn's satellite Rhea is a homogeneous mix of rock and ice, 
Geophys. Res. Lett. 34, L02202

\bibitem{and1} Anderson, J. D., Schubert, G., Jacobsen, R. A., Lau, E. L., Moore, W. B., Sjogren, W. L., 1998. 
Distribution of Rock, Metals, and Ices in Callisto, Science 280, 1573-1576

\bibitem[Asphaug and Melosh(1993)]{1993Icar..101..144A} Asphaug, E., 
Melosh, H.~J.\ 1993.\ The Stickney impact of PHOBOS - A dynamical model.\ 
Icarus 101, 144-164. 

\bibitem[Bai and Stone(2010)]{2010ApJ...722.1437B} Bai, X.-N., Stone, 
J.~M.\ 2010.\ Dynamics of Solids in the Midplane of Protoplanetary Disks: 
Implications for Planetesimal Formation.\ The Astrophysical Journal 722, 
1437-1459. 

\bibitem{barr2} Barr, A. C., 2008. Mobile Lid Convection Beneath Enceladus' South Polar Terrain. J. Geophys. 
Res. 113, E07009

\bibitem{marr3} Barr, A. C., McKinnon, W. B., 2007. Convection in Enceladus ice shell: Conditions for initiation. 
Geophys. Res. Lett. 34, L09202 

\bibitem{barr} Barr, A. C., Canup, R. M. 2008. Constraints on gas giant satellite formation from the interior 
states of partially diﬀerentiated satellites. Icarus 198, 163-177

\bibitem{tempel_spin}
Belton, M. J., Drahus, M. 2007.
The Accelerating Spin Of 9P/Tempel 1.
American Astronomical Society DPS meeting \#39, abstract \#43.05; BAAS 39, 498

\bibitem{belton_DPS}
Belton, M. J. S., Melosh, H. J. 2008.
Fluidization and Multiphase Transport of Particulate Cometary Material as an Explanation of the Smooth Terrains on 9P/Tempel 1. 
American Astronomical Society, DPS meeting \#40, abstract \#2.05

\bibitem{belton_icarus}
Belton, M.J. S., Feldman, P.D., A'Hearn, M.F., Carcich, B. 2008.
Cometary cryo-volcanism: Source regions and a model for the UT 2005 June 14 and other mini-outbursts on Comet 9P/Tempel 1.
Icarus 198, 189-207 

\bibitem[Benner et 
al.(2002)]{2002M&PS...37..779B} Benner, L.~A.~M., and 10 colleagues 2002.\ Radar observations of asteroid 1999 JM8.\ 
Meteoritics and Planetary Science 37, 779-792. 

\bibitem[Benz and Asphaug(1999)]{1999Icar..142....5B} Benz, W., Asphaug, 
E.\ 1999.\ Catastrophic Disruptions Revisited.\ Icarus 142, 5-20. 

\bibitem{binzel2010}
Binzel, R.P., A. Morbidelli, S. Merouane, F.E. DeMeo, M. Birlan, P. Vernazza, C.A. Thomas, A.S. Rivkin, S.J. Bus, A.T. Tokunaga 2010.
Earth encounters as the origin of fresh surfaces on near-Earth asteroids. Nature 463, 331-334

\bibitem[Bottke et al.(2001)]{2001Sci...294.1693B} Bottke, W.~F., 
Vokrouhlick{\'y}, D., Broz, M., Nesvorn{\'y}, D., Morbidelli, A.\ 2001.\ 
Dynamical Spreading of Asteroid Families by the Yarkovsky Effect.\ Science 
294, 1693-1696

\bibitem[Bottke et al.(2005)]{2005Icar..175..111B} Bottke, W.~F., Durda, 
D.~D., Nesvorn{\'y}, D., Jedicke, R., Morbidelli, A., Vokrouhlick{\'y}, D., 
Levison, H.\ 2005.\ The fossilized size distribution of the main asteroid 
belt.\ Icarus 175, 111-140. 

\bibitem{breiter_itokawa}
Breiter, S., P. Bartczak, M. Czekaj, B. Oczujda, D. Vokrouhlick\'y 2009.
The YORP effect on 25 143 Itokawa. Astronomy \& Astrophysics 507, 1073-1081

\bibitem{cas1} Castillo-Rogez, J., McCord, T. B., 2010. Ceres' evolution and present state constrained by shape data. 
Icarus 205, 443 

\bibitem{cas2} Castillo-Rogez, J. C., Matson, D. L., Sotin, C., Johnson, T. V., Lunine, J. I., Thomas, P. C., 2007. 
Iapetus' geophysics: Rotation rate, shape, and equatorial ridge. Icarus 190, 179-202

\bibitem{CMDA_cicalo}
Cical\`o, S., Scheeres, D.J. 2010.  Averaged rotational dynamics of an asteroid
in tumbling rotation under the YORP torque. Celestial Mechanics and Dynamical Astronomy 106, 301-337

\bibitem{colwell}
Colwell, J. E., A.A. Gulbis, M. Horanyi, S. Robertson 2005. Dust Transport in Photoelectron Layers
and the Formation of Dust Ponds on Eros. Icarus 175, 159-169

\bibitem{crifo}
Crifo J. F., Rodionov, A. V. 1997.
The Dependence of the Circumnuclear Coma Structure on the Properties of the Nucleus.
Icarus 127, 319-353

\bibitem{cuk2} \'Cuk, M., 2007. Formation and Destruction of Small Binary Asteroids.
ApJ 659, L57-L60.

\bibitem{cuk} \'Cuk, M., Burns, J. A., 2005. Effects of thermal radiation on the dynamics of binary NEAs.
Icarus 176, 418-431

\bibitem{cuk_nes}
\'Cuk, M., Nesvorn\'y, D. 2010.  Orbital evolution of small binary asteroids.
Icarus 207, 732-743

\bibitem[Cuzzi et al.(1993)]{1993Icar..106..102C} Cuzzi, J.~N., 
Dobrovolskis, A.~R., Champney, J.~M.\ 1993.\ Particle-gas dynamics in the 
midplane of a protoplanetary nebula.\ Icarus 106, 102

\bibitem[Cuzzi et al.(2001)]{2001ApJ...546..496C} Cuzzi, J.~N., Hogan, 
R.~C., Paque, J.~M., Dobrovolskis, A.~R.\ 2001.\ Size-selective 
Concentration of Chondrules and Other Small Particles in Protoplanetary 
Nebula Turbulence.\ The Astrophysical Journal 546, 496-508

\bibitem[Cuzzi et al.(2008)]{2008ApJ...687.1432C} Cuzzi, J.~N., Hogan, 
R.~C., Shariff, K.\ 2008.\ Toward Planetesimals: Dense Chondrule Clumps in 
the Protoplanetary Nebula.\ The Astrophysical Journal 687, 1432-1447

\bibitem[Cuzzi et al.(2010)]{2010Icar..208..518C} Cuzzi, J.~N., Hogan, 
R.~C., Bottke, W.~F.\ 2010.\ Towards initial mass functions for asteroids 
and Kuiper Belt Objects.\ Icarus 208, 518-538 

\bibitem{dod1} Dodson-Robinson, S. E., Bodenheimer, P., 2010. The Formation of Uranus and Neptune in Solid-Rich 
Feeding Zones: Connecting Chemistry and Dynamics. Icarus 207, 491

\bibitem{dod2} Dodson-Robinson, S. E., Willacy, K., Bodenheimer, P., Turner, N. J., Beichman, C., 2009. 
Ice Lines, Planetesimal Composition and Solid Surface Density in the Solar Nebula. Icarus 200, 672

\bibitem{dur} Durda, D. D.,  Bottke, W. F.,  Enke, B. L.,  et al., 2004.
The formation of asteroid satellites in large impacts: results from numerical simulations.
Icarus 170, 243-257

\bibitem[Durda et al.(2007)]{2007Icar..186..498D} Durda, D.~D., Bottke, 
W.~F., Nesvorn{\'y}, D., Enke, B.~L., Merline, W.~J., Asphaug, E., 
Richardson, D.~C.\ 2007.\ Size-frequency distributions of fragments from 
SPH/N-body simulations of asteroid impacts: Comparison with observed 
asteroid families.\ Icarus 186, 498-516

\bibitem{elk} Elkins-Tanton, L., Weiss, B. P., Zuber, M. T., 2010. Chondrites as samples of differentiated 
planetesimals. DPS meeting, abstract 2.04

\bibitem{Icarus_YORPexpansion}
Fahnestock, E. G., Scheeres, D. J. 2009.
Binary asteroid orbit expansion due to continued YORP spin-up of the primary
and primary surface particle motion. Icarus 201, 135-152

\bibitem{goguen_LPSC}
Goguen, J. D., Thomas, P. C., Veverka, J. F. 2008.
Flows on the Nucleus of Comet Tempel 1.
39th Lunar and Planetary Science Conference, LPI Contribution No. 1391, 1969

\bibitem[Goldreich and Ward(1973)]{1973ApJ...183.1051G} Goldreich, P., 
Ward, W.~R.\ 1973.\ The Formation of Planetesimals.\ The Astrophysical 
Journal 183, 1051-1062 

\bibitem{goldreich}
Goldreich, P., Sari, R. 2009. Tidal evolution of rubble piles. ApJ 691, 54-60

\bibitem{gos} Gosh, A., McSween, H. Y., 1998. Thermal Model for the Differentiation of Asteroid 4 Vesta, 
Based on Radiogenic Heating. Icarus 134, 187-206

\bibitem{gri} Grimm, R.E., McSween, H.Y., 1989. Water and the thermal evolution of carbonaceous chondrite 
parent bodies. Icarus 82, 244-280

\bibitem{harris_shape}
Harris, A. W., Fahnestock, E. G., Pravec, P. 2009.
On the shapes and spins of rubble pile asteroids. Icarus 199, 310-318

\bibitem{holsappleA} 
Holsapple, K. A. 2001.  Equilibrium Configurations of Solid Cohesionless Bodies. Icarus 154,  432-448

\bibitem{holsappleB} 
Holsapple, K. A.  2004.  Equilibrium figures of spinning bodies with self-gravity. Icarus 172, 272-303

\bibitem{holsapple_fast}
Holsapple, K. A. 2007.
Spin limits of Solar System bodies: From the small fast-rotators to 2003 EL61. Icarus 187, 500-509

\bibitem{holsapple_YORP}
Holsapple, K. A.  2010. On YORP-induced spin deformations of asteroids. Icarus 205, 430-442

\bibitem[Housen and Holsapple(2003)]{2003Icar..163..102H} Housen, K.~R., 
Holsapple, K.~A.\ 2003.\ Impact cratering on porous asteroids.\ Icarus 163, 
102-119. 

\bibitem{hub} Hubbard, W. B., Podolak, M., Stevenson, D. J., 1995, The Interior of Neptune. In Neptune and Triton.
109

\bibitem{hus} Hussmann, H., Sohl, F., Spohn, T., 2006. Subsurface oceans and deep interiors of medium-sized outer planet 
satellites and large trans-neptunian objects. Icarus 185, 258

\bibitem{iess} Iess, L., N. J. Rappaport, P. Tortora, J. I. Lunine, J. W. Armstrong, S. W. Asmar, L. Somenzi, 
F. Zingoni, 2007. Gravity and interior of Rhea from Cassini data analysis, Icarus 190, 585-593

\bibitem{DPS_09_seth}
Jacobson, S. A., Scheeres, D. J. 2009.  A Rapid Phase of Tidal Dissipation for Post-Fission Binary Asteroids. 
41st Annual American Astronomical Society DPS meeting, abstract \#56.08

\bibitem{DDA_10_seth}
Jacobson, S. A., Scheeres, D. J. 2010. Formation of Observed Asteroid Systems by Rotational Fission. 
American Astronomical Society DDA meeting

\bibitem{jewitt_comets}
Jewitt, D. C. 2004. From Cradle to Grave:  The Rise and Demise of the Comets. In 
Comets II, M. C. Festou, H. U. Keller, and H. A. Weaver (eds.), University of Arizona Press, Tucson, 659-676

\bibitem[Jewitt et al.(2010)]{2010Natur.467..817J} Jewitt, D., Weaver, H., 
Agarwal, J., Mutchler, M., Drahus, M.\ 2010.\ A recent disruption of the 
main-belt asteroid P/2010A2.\ Nature 467, 817-819. 

\bibitem[Johansen and Youdin(2007)]{2007ApJ...662..627J} Johansen, A., 
Youdin, A.\ 2007.\ Protoplanetary Disk Turbulence Driven by the Streaming 
Instability: Nonlinear Saturation and Particle Concentration.\ The 
Astrophysical Journal 662, 627-641 

\bibitem[Johansen et al.(2009)]{2009ApJ...704L..75J} Johansen, A., Youdin, 
A., Mac Low, M.-M.\ 2009.\ Particle Clumping and Planetesimal Formation 
Depend Strongly on Metallicity.\ The Astrophysical Journal 704, L75-L79

\bibitem[Johansen et al.(2007)]{2007Natur.448.1022J} Johansen, A., Oishi, 
J.~S., Mac Low, M.-M., Klahr, H., Henning, T., Youdin, A.\ 2007.\ Rapid 
planetesimal formation in turbulent circumstellar disks.\ Nature 448, 
1022-1025

\bibitem[Jutzi et al.(2008)]{2008Icar..198..242J} Jutzi, M., Benz, W., 
Michel, P.\ 2008.\ Numerical simulations of impacts involving porous 
bodies. I. Implementing sub-resolution porosity in a 3D SPH hydrocode.\ 
Icarus 198, 242-255. 

\bibitem{kas} Kaasalainen, M.,  Durech, J.,  Warner, B.D.,  et al., 2007.
Acceleration of the rotation of asteroid 1862 Apollo by radiation torques.
Nature 446, 420-422

\bibitem{kar} Kargel, J. S.1991. Brine Volcanism and the Interior Structures of 
Asteroids and Icy Satellites, Icarus 94, 368-390.

\bibitem{kie} Kieffer, S., Lu, X., Bethke, C. M., Spencer, J. R., Marshak, S., Navrotsky, A., 2006. A clathrate 
reservoir hypothesis for Enceladus' south polar plume. Science 314, 1764-1766

\bibitem{tempel_II}
Knight, M. M., T.L. Farnham, D.G. Schleicher, E.W. Schwieterman 2010.
The Increasing Rotation Period of Comet 10P/Tempel 2. arXiv:1009.3019v1 

\bibitem[Korycansky and Asphaug(2006)]{2006Icar..181..605K} Korycansky, 
D.~G., Asphaug, E.\ 2006.\ Low-speed impacts between rubble piles modeled 
as collections of polyhedra.\ Icarus 181, 605-617. 

\bibitem{lee2}
Lee, P. 1996. Dust Levitation on Asteroids. Icarus 124, 181-194

\bibitem[Leinhardt and Stewart(2009)]{2009Icar..199..542L} Leinhardt, 
Z.~M., Stewart, S.~T.\ 2009.\ Full numerical simulations of catastrophic 
small body collisions.\ Icarus 199, 542-559. 

\bibitem[Levison et al.(2008)]{2008Icar..196..258L} Levison, H.~F., 
Morbidelli, A., Vanlaerhoven, C., Gomes, R., Tsiganis, K.\ 2008.\ Origin of 
the structure of the Kuiper belt during a dynamical instability in the 
orbits of Uranus and Neptune.\ Icarus 196, 258-273. 

\bibitem{lop} Lopes, R. M. C., and 43 coauthors, 2007. Cryovolcanic features on Titan's surface as revealed by 
the Cassini Titan Radar Mapper. Icarus 186, 395

\bibitem{low} Lowry, S. C.,  Fitzsimmons, A.,  Pravec, P.,  et al., 2007.
Direct Detection of the Asteroidal YORP Effect. Science 316, 272

\bibitem{mccor}McCord, T.B., Sotin, C., 2005. Ceres: Evolution and current state. J. Geophys. Res. 110.
CiteID E05009.

\bibitem[McMahon and Scheeres(2010)]{2010Icar..209..494M} McMahon, J., 
Scheeres, D.\ 2010.\ Detailed prediction for the BYORP effect on binary 
near-Earth Asteroid (66391) 1999 KW4 and implications for the binary 
population.\ Icarus 209, 494-509. 

\bibitem[Michel et al.(2001)]{2001Sci...294.1696M} Michel, P., Benz, W., 
Tanga, P., Richardson, D.~C.\ 2001.\ Collisions and Gravitational 
Reaccumulation: Forming Asteroid Families and Satellites.\ Science 294, 
1696-1700. 

\bibitem[Michel et al.(2002)]{2002Icar..160...10M} Michel, P., Tanga, P., 
Benz, W., Richardson, D.~C.\ 2002.\ Formation of Asteroid Families by 
Catastrophic Disruption: Simulations with Fragmentation and Gravitational 
Reaccumulation.\ Icarus 160, 10-23. 

\bibitem[Michel et al.(2003)]{2003Natur.421..608M} Michel, P., Benz, W., 
Richardson, D.~C.\ 2003.\ Disruption of fragmented parent bodies as the 
origin of asteroid families.\ Nature 421, 608-611. 

\bibitem{MM2009}
Minton, D. A., Malhotra, R. 2009.
A record of planet migration in the main asteroid belt. Nature 457, 1109-1111

\bibitem[Morbidelli et al.(2009)]{2009Icar..204..558M} Morbidelli, A., 
Bottke, W.~F., Nesvorn{\'y}, D., Levison, H.~F.\ 2009.\ Asteroids were born 
big.\ Icarus 204, 558-573

\bibitem{mysen}
Mysen, E. 2008. An analytical model for YORP and Yarkovsky effects with a physical thermal lag. Astronomy and 
Astrophysics 484,  563-573

\bibitem{neishtadt_icarus}
Neishtadt, A. I, {D. J. Scheeres} , V. V. Sidorenko, A. A. Vasiliev. 2002.
Evolution of comet nucleus rotation. Icarus 157, 205-218

\bibitem[Nesvorn{\'y} and Vokrouhlick{\'y}(2007)]{2007AJ....134.1750N} 
Nesvorn{\'y}, D., Vokrouhlick{\'y}, D.\ 2007.\ Analytic Theory of the YORP 
Effect for Near-Spherical Objects.\ The Astronomical Journal 134, 1750. 

\bibitem[Nesvorn{\'y} et al.(2005)]{2005Icar..173..132N} Nesvorn{\'y}, D., 
Jedicke, R., Whiteley, R.~J., Ivezi{\'c}, {\v Z}.\ 2005.\ Evidence for 
asteroid space weathering from the Sloan Digital Sky Survey.\ Icarus 173, 
132-152. 

\bibitem[Nesvorn{\'y} et al.(2006)]{2006Icar..183..296N} Nesvorn{\'y}, D., 
Enke, B.~L., Bottke, W.~F., Durda, D.~D., Asphaug, E., Richardson, D.~C.\ 
2006.\ Karin cluster formation by asteroid impact.\ Icarus 183, 296-311

\bibitem[Nesvorn{\'y} et al.(2010)]{2010AJ....140..785N} Nesvorn{\'y}, D., 
Youdin, A.~N., Richardson, D.~C.\ 2010a.\ Formation of Kuiper Belt Binaries 
by Gravitational Collapse.\ The Astronomical Journal 140, 785-793 

\bibitem{nesvorny2010}
Nesvorn\'y, D., Bottke, W. F., Vokrouhlick\'y, D., Chapman, C. R., Rafkin, S. 2010b.
Do planetary encounters reset surfaces of near Earth asteroids? Icarus 209, 510-519

\bibitem[Nesvorny et al.(2011)]{2011arXiv1102.5706N} Nesvorny, D., 
Vokrouhlicky, D., Bottke, W.~F., Noll, K., Levison, H.~F.\ 2011.\ Observed 
Binary Fraction Sets Limits on the Extent of Collisional Grinding in the 
Kuiper Belt.\ ArXiv e-prints arXiv:1102.5706. 

\bibitem{nim} Nimmo, F., Spencer, J. R., Pappalardo, R. T., Mullen, M. E., 2007. 
Shear heating as the origin of plumes and heat flux on Enceladus. Nature 447, 289-291

\bibitem{ost} Ostro, S. J.,  Margot, J.-L.,  Benner, L. A. M.,  et al., 2006.
Radar Imaging of Binary Near-Earth Asteroid (66391) 1999 KW4.
Science 314, 1276-1280

\bibitem[Pan and Sari(2005)]{2005Icar..173..342P} Pan, M., Sari, R.\ 2005.\ 
Shaping the Kuiper belt size distribution by shattering large but 
strengthless bodies.\ Icarus 173, 342-348. 

\bibitem[Parker and Kavelaars(2010)]{2010ApJ...722L.204P} Parker, A.~H., 
Kavelaars, J.~J.\ 2010.\ Destruction of Binary Minor Planets During Neptune 
Scattering.\ The Astrophysical Journal 722, L204-L208. 

\bibitem[Petit et al.(2002)]{2002aste.conf..711P} Petit, J.-M., Chambers, 
J., Franklin, F., Nagasawa, M.\ 2002.\ Primordial Excitation and Depletion 
of the Main Belt.\ Asteroids III 711-723. 

\bibitem{por} Porco, C. C., and coauthors, 2006. Cassini observes the active south pole of Enceladus. 
Science 311, 1393-1401

\bibitem{pravec_nature}
Pravec, P., D. Vokrouhlick\'y, D. Polishook, {D.J. Scheeres}, A. W. Harris, A. Galad, O.
Vaduvescu, F. Pozo, A. Barr, P. Longa, F. Vachier, F. Colas, D. P. Pray, J. Pollock, D. Reichart, K.
Ivarsen, J. Haislip, A. LaCluyze, P. Kusnirak, T. Henych, F. Marchis, B. Macomber,{S. A.
Jacobson}, Y. N. Krugly, A. Sergeev, and A. Leroy.  2010.  Formation of asteroid pairs by rotational 
fission. Nature 466, 1085-1088

\bibitem{pri} Prialnik, D., 2000. Modeling the Comet Nucleus Interior: Application to Comet C/1995 O1 Hale-Bopp. 
Earth, Moon, and Planets 89, 27-52

\bibitem{pri2} Prieto-Ballesteros, O., Kargel, J.S., 2005. Thermal state and complex geology of a
heterogeneous salty crust of Jupiter's satellite, Europa. Icarus 173, 212-221.

\bibitem{ray} Rayman, M. D., Fraschetti, T. C., Raymond, C. A., Russell, C. T., 2006. 
Dawn: A mission in development for exploration of main belt asteroids Vesta and Ceres. 
Acta Astronautica 58, 605

\bibitem[Richardson et al.(2002)]{2002aste.conf..501R} Richardson, D.~C., 
Leinhardt, Z.~M., Melosh, H.~J., Bottke, W.~F., Jr., Asphaug, E.\ 2002.\ 
Gravitational Aggregates: Evidence and Evolution.\ Asteroids III 501-515. 

\bibitem{Richardson2005}
Richardson, D. C., Elankumaran, P., Sanderson, R. E. 2005.
Numerical experiments with rubble piles: equilibrium shapes and spins.
Icarus 173, 349-361.

\bibitem{rob} Roberts, J. H., Nimmo, F., 2008. Tidal heating and the long-term stability of a subsurface ocean 
on Enceladus. Icarus 194, 675-689

\bibitem{robinson}
Robinson, M. S., P. C. Thomas, J. Veverka, S. Murchie, B. Carcich 2001.
The nature of ponded deposits on Eros. Nature 413, 396-400

\bibitem{DPS_09_rossi}
Rossi, A., Marzari, F., Scheeres, D. J. 2009.  Spin Evolution of Small Main Belt Asteroids.
41st American Astronomical Society DPS meeting, Puerto Rico,  abstract \#56.01

\bibitem{DPS_10_rossi}
Rossi, A., Marzari, F., Scheeres, D. J. 2010. Unveiling the excess of slow rotators in the small 
main belt asteroids. 42nd American Astronomical Society DPS meeting, Pasadena

\bibitem{Ru00}
Rubincam, D. P. 2000. Radiative Spin-up and Spin-down of small asteroids, Icarus 148, 2-11

\bibitem[Safronov(1969)]{1969QB981.S26......} Safronov, V.~S.\ 1969.\ 
Evoliutsiia doplanetnogo oblaka.\ Moscow: Nakua

\bibitem[S{\'a}nchez and Scheeres(2011)]{2011ApJ...727..120S} S{\'a}nchez, 
P., Scheeres, D.~J.\ 2011.\ Simulating Asteroid Rubble Piles with A 
Self-gravitating Soft-sphere Distinct Element Method Model.\ The 
Astrophysical Journal 727, 120. 

\bibitem{s4} Scheeres, D. J., 2007.
Rotational fission of contact binary asteroids.
Icarus 189, 370-385

\bibitem{PSS_fission}
Scheeres, D. J.  2009.  Minimum energy asteroid reconfigurations 
and catastrophic disruptions. Planetary and Space Science 57,  154-164

\bibitem{Itokawa_CM}
Scheeres, D. J., Gaskell, R. W.  2008.  Effect of density inhomogeneity on YORP:  
The case of Itokawa. Icarus 198, 125-129

\bibitem[Scheeres and Mirrahimi(2008)]{2008CeMDA.101...69S} Scheeres, 
D.~J., Mirrahimi, S.\ 2008.\ Rotational dynamics of a solar system body 
under solar radiation torques.\ Celestial Mechanics and Dynamical Astronomy 
101, 69-103

\bibitem{KW4_scheeres}
Scheeres, D. J., 
{E. G. Fahnestock}, 
S. J. Ostro, 
J.-L. Margot, 
L. A. M. Benner, 
{S. B. Broschart}, 
{J. Bellerose}, 
J. D. Giorgini, 
M. C. Nolan, 
C. Magri, 
P. Pravec, 
P. Scheirich, 
R. Rose, 
R. F. Jurgens, 
S. Suzuki, 
E. M. DeJong  2006. Dynamical Configuration of Binary Near-Earth Asteroid (66391) 1999 KW4. 
Science 314, 1280-1283

\bibitem{s2} Scheeres, D. J., Fahnestock, E. G., Ostro, S. J., et al., 2006. 
Dynamical Configuration of Binary Near-Earth Asteroid (66391) 1999 KW4.
Science 314, 1280-1283

\bibitem{itokawa_yorp}
Scheeres, D. J., M. Abe, M. Yoshikawa, R. Nakamura, R. W. Gaskell, P.A. Abell  2007.  
The effect of YORP on Itokawa. Icarus 188, 425-429

\bibitem{Icarus_cohesion}
Scheeres, D. J., Hartzell, C. M., S\`anchez, P., Swift, M.  2010.  
Scaling forces to asteroid surfaces: The role of cohesion. Icarus 210, 968-984

\bibitem[Schlichting and Sari(2011)]{2011ApJ...728...68S} Schlichting, 
H.~E., Sari, R.\ 2011.\ Runaway Growth During Planet Formation: Explaining 
the Size Distribution of Large Kuiper Belt Objects.\ The Astrophysical 
Journal 728, 68. 

\bibitem{schu} Schubert, G., J. D. Anderson, B. J. Travis, J. Palguta, 2007. Enceladus: Present internal structure 
and diferentiation by early and long-term radiogenic heating. Icarus 188, 345-355

\bibitem[Sekiya(1998)]{1998Icar..133..298S} Sekiya, M.\ 1998.\ 
Quasi-Equilibrium Density Distributions of Small Dust Aggregations in the 
Solar Nebula.\ Icarus 133, 298-309. 

\bibitem[Snodgrass et al.(2010)]{2010Natur.467..814S} Snodgrass, C., and 18 
colleagues 2010.\ A collision in 2009 as the origin of the debris trail of 
asteroid P/2010A2.\ Nature 467, 814-816. 

\bibitem{spo} Spohn, T., Schubert, G., 2003. Oceans in the icy Galilean satellites 
of Jupiter? Icarus 161, 456

\bibitem{statler_YORP}
Statler, T. S. 2009. Extreme sensitivity of the YORP effect to small-scale topography. Icarus 202, 502-513

\bibitem{tay} Taylor, P. A.,  Margot, J.-L.,  Vokrouhlick\'y, D.,  et al., 2007.
Spin Rate of Asteroid (54509) 2000 PH5 Increasing Due to the YORP Effect.
Science 316, 274

\bibitem{tho324} Thomas, C., Parker, J.Wm., McFadden, L.A., Russell, C.T., Stern, S.A., Sykes, M.V.,
Young, E.F., 2005. Differentiation of the asteroid Ceres as revealed by its shape.
Nature 437, 224-226

\bibitem{tobie} Tobie, G., Lunine, J. I., and Sotin, C., 2006.  Episodic outgassing as the origin of atmospheric methane 
on Titan.  Nature 440, 61-64

\bibitem{tra} Travis, B. E., Schubert, G., 2005. Hydrothermal convection in carbonaceous chondrite parent 
bodies. Earth and Planetary Science Letters 240, 234-250

\bibitem{VF2000}
Vokrouhlick\'y, D., Farinella, P. 2000. 
Efficient delivery of meteorites to the Earth from a wide range of asteroid parent bodies. Nature 407, 606-608

\bibitem{VoCa02} 
Vokrouhlick\'y, D., \v{C}apek, D. 2002.
YORP-Induced long-term evolution of the spin state of small asteroids and meteoroids:  
Rubincam's Approximation. Icarus 159, 449-467

\bibitem{vok} Vokrouhlick\'y, D., Nesvorn\'y, D. 2008. Pairs of Asteroids Probably of a 
Common Origin. AJ 136, 280-290

\bibitem{VoNeBo03} 
Vokrouhlick\'y, D., Nesvorn\'y, D., Bottke, W. F. 2003.
The vector alignments of asteroid spins by thermal torques. Nature 425, 147-151

\bibitem{Icarus_tumblingYORP}
Vokrouhlick\'y, D., Breiter, S., Nesvorn\'y, D., Bottke, W. F.  2007.
Generalized YORP evolution: Onset of tumbling and new asymptotic states.
Icarus 191, 636-650

\bibitem{walsh}
Walsh, K. J., Richardson, D. C., Michel, P. 2008.  Rotational breakup as the origin of 
small binary asteroids. Nature 454, 188-191

\bibitem[Weidenschilling(1984)]{1984Icar...60..553W} Weidenschilling, 
S.~J.\ 1984.\ Evolution of grains in a turbulent solar nebula.\ Icarus 60, 
553-567

\bibitem{wei} Weiss, B., Caporzen, L., Elkins-Tanton, L., Shuster, D. L., Ebel, D. S., Gattacceca, J., 
Binzel, R. P., 2010. DPS meeting, abstract 2.05

\bibitem{wur} Wurz, P., and 13 coauthors, 2010. Possible detection of water in the exosphere of (21) Lutetia. 
American Geophysical Union Fall Meeting, abstract P14B-03

\bibitem[Youdin and Shu(2002)]{2002ApJ...580..494Y} Youdin, A.~N., Shu, 
F.~H.\ 2002.\ Planetesimal Formation by Gravitational Instability.\ The 
Astrophysical Journal 580, 494-505

\bibitem[Youdin and Goodman(2005)]{2005ApJ...620..459Y} Youdin, A.~N., 
Goodman, J.\ 2005.\ Streaming Instabilities in Protoplanetary Disks.\ The 
Astrophysical Journal 620, 459-469
 
\end{thebibliography}
\end{document}